\documentclass[a4paper,11pt]{article}
\usepackage{pos}
\usepackage{multirow}

\title{Overview of the physics prospects in MOMENT}

\author*{Sampsa Vihonen}

\affiliation{School of Physics, Sun Yat-sen University,\\
  No. 135 Xingang Xi Road, 510275 Guangzhou, P.R. China}

\emailAdd{sampsa@mail.sysu.edu.cn}

\abstract{MuOn-decay MEdium-baseline NeuTrino beam facility (MOMENT) is a recently proposed neutrino oscillation experiment that is currently under consideration in China. Based on a novel accelerator concept, MOMENT is capable of delivering 15~MW neutrino beam produced with the decay of positively and negatively charged muons. In this study, the physics performance of MOMENT is briefly reviewed in its baseline setup, involving a 150~km baseline length and large gadolinium-doped Water Cherenkov detector. The prospects are discussed in the case of precision measurements on the Dirac {\em CP} phase $\delta_{CP}$, which MOMENT is shown to be able to measure by about 8$^\circ$...18$^\circ$ resolution at 1$\,\sigma$ CL. It is examined how MOMENT performs as an independent experiment and also as a complementary facility to the future superbeam experiments.}

\FullConference{%
  *** The 22nd International Workshop on Neutrinos from Accelerators (NuFact2021) ***\\
  *** 6--11 Sep 2021 ***\\
  *** Cagliari, Italy ***}

\begin{document}
\maketitle

\section{Introduction}
\label{sec:intro}

\noindent Recent advances in neutrino oscillation physics have led to a significant improvement in the understanding of three-neutrino mixing~\cite{Esteban:2020cvm}. The neutrino mixing angles $\theta_{12}$, $\theta_{13}$ and $\theta_{23}$ and mass-squared differences $\Delta m_{21}^2$ and $\Delta m_{31}^2$ have been measured in the numerous solar, reactor and accelerator neutrino experiments as well as in neutrino telescopes. The next-generation neutrino oscillation experiments such as the Jiangmen Underground Neutrino Observatory (JUNO), Deep Underground Neutrino Experiment (DUNE) and Tokai-to-HyperKamiokande (T2HK) experiment will aim to address the remaining questions in the standard neutrino oscillation parameters\footnote{One of the long-term goals in neutrino experiments is to measure the value of the Dirac {\em CP} phase $\delta_{CP}$, which could shed light on the problem of matter-antimatter asymmetry as well as the origin of neutrino mixing.}. Future experiments are also expected to improve the precision on the previously measured oscillation parameters and look for signals of new physics.

MuOn-decay MEdium-baseline NeuTrino beam facility (MOMENT) is a recently proposed neutrino oscillation experiment currently under consideration in China~\cite{Cao:2014bea}. MOMENT features a novel accelerator concept where a high-intensity low-energy beam of neutrinos and antineutrinos is produced by using a continuous-wave proton beam and a stack of super-conducting solenoids. The neutrino beams in MOMENT are created by using positively and negatively charged muons, which allow to measure neutrino mixing in eight oscillation channels in low beam-related backgrounds. This approach has many advantages, including low beam-related backgrounds and greater access to oscillation channels. In this proceeding, we briefly review the physics prospects of MOMENT in the precision measurement of the standard neutrino oscillation parameters. We focus on the sensitivity to Dirac {\em CP} phase $\delta_{CP}$. Potential synergies with the Chinese research planning are also discussed. We also highlight the complementarity between MOMENT and the proposed next-generation long-baseline neutrino experiments DUNE, T2HK and T2HKK.

This proceeding is organized as follows. We give a brief overview on the MOMENT configuration in section\,\ref{sec:overview} and discuss the prospects to measure $\delta_{CP}$ in section\,\ref{sec:results}. We finally leave the concluding remarks in section\,\ref{sec:concl}.

\section{Overview of MOMENT}
\label{sec:overview}

\begin{figure}[!ht]
        \center{\includegraphics[width=0.75\textwidth]{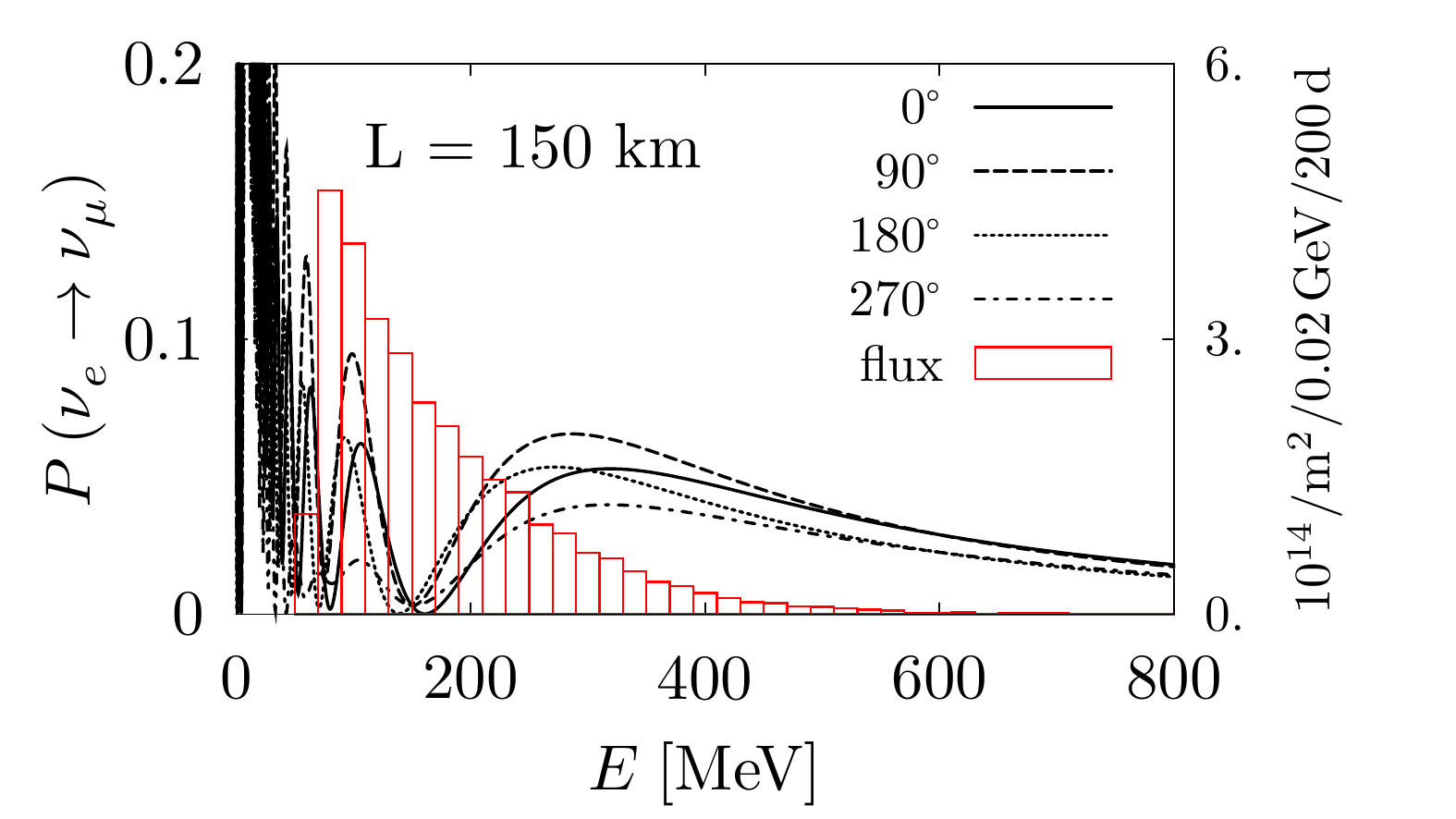}}
        \caption{Transition probabilities for the oscillation channel $\nu_e \rightarrow \nu_\mu$ for neutrino energies and baseline length relevant for the MOMENT setup~\cite{Tang:2019wsv}. The probabilities are shown for $\delta_{CP} =$ 0$^\circ$, 90$^\circ$, 180$^\circ$ and 270$^\circ$, as indicated by the solid, dashed, dotted and dot-dashed curves, respectively. The expected $\nu_e$ fluxes in MOMENT are represented by the red boxes.}
        \label{fig:fluxes}
\end{figure}

\noindent MOMENT is a next-generation muon-decay-based neutrino beam facility which utilizes a novel technology to generate a low-energy high-intensity beam~\cite{Cao:2014bea}. Following the design and timeline of the China initiative for Accelerator-Driven Systems (CiADS), the proton beam of MOMENT is expected to reach 15~MW power at 1.5~GeV energy. The neutrino beam of MOMENT reaches its peak at about 200-300~MeV and it is therefore suitable for a baseline length of 150~km. The synergy of this configuration can be seen in Figure~\ref{fig:fluxes}, where the simulated $\nu_e$ fluxes and oscillation probabilities are shown. The expected neutrino fluxes can easily reach the second oscillation maximum and parts of the first maximum.

As a muon-decay-based neutrino beam facility, MOMENT has many advantages known from the conventional neutrino factory, such as the access to eight different oscillation channels and a well-understood neutrino source. MOMENT is anticipated to have excellent control of beam-related systematic uncertainties and low beam-induced backgrounds. The average energy of the neutrino beam in MOMENT places neutrino interactions in the quasi-elastic regime, which requires a large-scale neutrino detector that is capable of studying neutrinos at the desired energy range. Based on the recent progress in the gadolinium-doped Water Cherenkov technology, a neutrino detector of 500~kt fiducial mass has been considered for the far detector of MOMENT~\cite{Cao:2014bea,Tang:2019wsv}. Alternative designs for the neutrino detector have also been examined~\cite{Tang:2020xoy}.

The prospects of the MOMENT facility have been studied in the precision measurement of the Dirac {\em CP} phase~\cite{Tang:2019wsv} and in various searches for new physics, see Ref.~\cite{Tang:2021lyn} for more information. MOMENT could be established by constructing entirely new facilities for source and detector. Alternatively, the configuration could be realized using the existing research infrastructure in China. Table~\ref{tab:labsites2} presents the geographical locations of five accelerator laboratories and two underground laboratories which are either currently under construction or are under active consideration. Placing MOMENT beamline at CiADS and detector at JUNO for instance would give access to 221~km baseline length.

\begin{table}[!ht]
\caption{\label{tab:labsites2}Geographical locations of the existing and proposed accelerator and underground laboratories~\cite{Tang:2021lyn}. The baseline lengths as well as the energies needed for the first and second oscillation maximum are shown for each potential configuration. The neutrino energies are shown for $\Delta m_{31}^2 \simeq$ 2.517$\times$10$^{-3}$~eV$^2$.}
\begin{center}
\resizebox{\linewidth}{!}{
\begin{tabular}{ |c|c|c|c|c|c|c| } 
 \hline
 \multirow{2}{*}{Accelerator facility} & \multicolumn{3}{ c| }{JUNO~(22.12$^\circ$,~112.51$^\circ$)} & \multicolumn{3}{ c| }{CJPL~(28.15$^\circ$,~101.71$^\circ$)} \\
 \cline{2-7}
 & Baseline & 1$^{\rm st}$ maximum & 2$^{\rm nd}$ maximum & Baseline & 1$^{\rm st}$ maximum & 2$^{\rm nd}$ maximum \\
 \hline
 CAS-IMP~(36.05$^\circ$,~103.68$^\circ$) & {1759~km} & {3.6~GeV} & {1.2~GeV} & {894~km} & {1.8~GeV} & {600~MeV} \\ \hline
 CiADS~(23.08$^\circ$,~114.40$^\circ$) & {221~km} & {450~MeV} & {150~MeV} & {1389~km} & {2.8~GeV} & {940~MeV} \\ \hline
 CSNS~(23.05$^\circ$,~113.73$^\circ$) & {162~km} & {330~MeV} & {110~MeV} & {1329~km} & {2.7~GeV} & {900~MeV} \\ \hline
 Nanjing~(32.05$^\circ$,~118.78$^\circ$) & {1261~km} & {2.6~GeV} & {850~MeV} & {1693~km} & {3.4~GeV} & {1.1~GeV} \\ \hline
 SPPC~(39.93$^\circ$,~116.40$^\circ$) & {1871~km} & {3.8~GeV} & {1.3~GeV} & {1736~km} & {3.5~GeV} & {1.2~GeV} \\
 \hline
\end{tabular}}
\end{center}
\end{table}

\section{Physics prospects}
\label{sec:results}

\begin{figure}[!ht]
        \center{\includegraphics[width=1.0\textwidth]{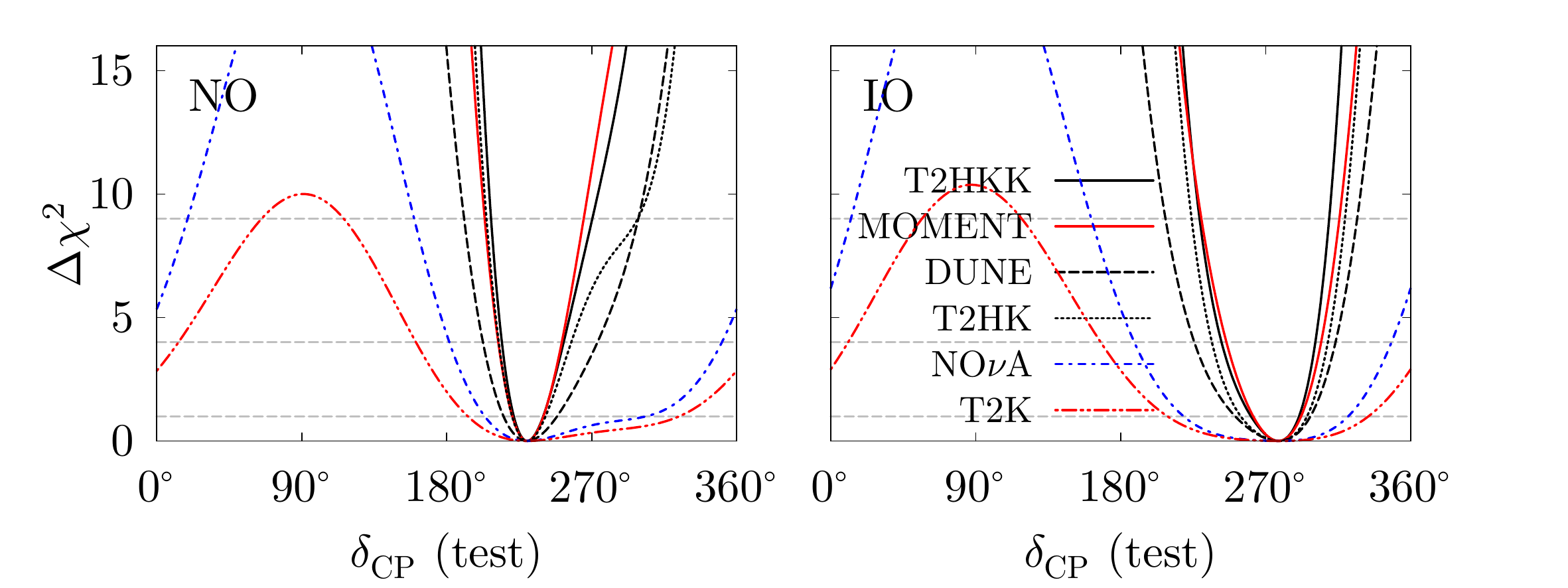}}
        \caption{Sensitivity to Dirac {\em CP} phase $\delta_{CP}$ in MOMENT and several long-baseline neutrino experiments. The true value is taken to be $\delta_{CP} \simeq$ 230$^\circ$ for normal ordering (NO) and 278$^\circ$ for inverted ordering (IO). The figure is adapted updated from the similar one in Ref.~\cite{Tang:2019wsv}.}
        \label{fig:DChi2}
\end{figure}

\noindent MOMENT has been shown to be very capable of performing precision measurements on the neutrino oscillation parameters. The sensitivity to the Dirac {\em CP} phase $\delta_{CP}$ in the baseline setup of MOMENT is comparable with the expected sensitivities that have been projected for the future long-baseline experiments~\cite{Tang:2019wsv}. This is demonstrated in Figure~\ref{fig:DChi2}, where the sensitivities to $\delta_{CP}$ are presented for the presently running experiments T2K and NO$\nu$A, future experiments DUNE and T2HK. Sensitivities are also shown for the proposed T2HKK configuration, where the second detector of T2HK is constructed in South Korea. The sensitivities are displayed separately for the normal ordering (NO) and inverted ordering (IO) of neutrino masses. The true values of the neutrino oscillation parameters are taken to be identical with the present fit result from the \texttt{nu-fit 5.1} database~\cite{Esteban:2020cvm}. In this regard, MOMENT is able to provide sensitivities that are comparable to those obtained for DUNE, T2HK and T2HKK.

\begin{figure}[!t]
        \center{\includegraphics[width=1.0\textwidth]{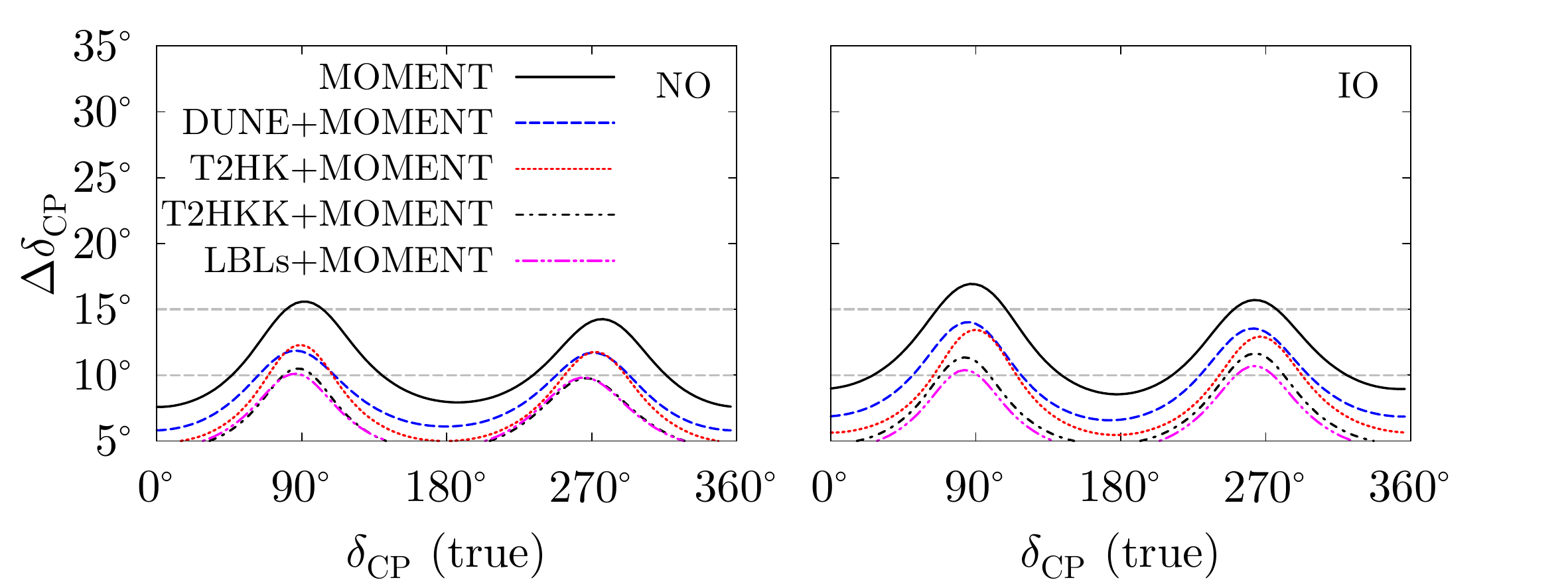}}
        \caption{Expected precisions on the {\em CP} phase $\delta_{CP}$ in MOMENT~\cite{Tang:2019wsv}. The combined sensitivities for MOMENT and the long-baseline experiments DUNE, T2HK and T2HKK are shown in terms of absolute precision at 1$\,\sigma$ CL. Here LBLs+MOMENT stands for the combined run of DUNE, T2HKK and MOMENT. Sensitivities are shown for normal ordering (NO) and inverted ordering (IO).}
        \label{fig:Ddelta}
\end{figure}

There exists synergy between MOMENT and the proposed long-baseline neutrino experiments. This complementarity is illustrated in Figure~\ref{fig:Ddelta}, where the expected resolution to $\delta_{CP}$ is projected in the proposed configurations of MOMENT, DUNE, T2HK and T2HKK. The figure shows the absolute precision for $\delta_{CP}$ as function of its theoretically allowed values~\footnote{It should be noted that at 3$\,\sigma$~CL, the recent experimental data disfavours values $\delta_{CP} =$ -10$^\circ$...144$^\circ$ in normal ordering and -15$^\circ$...194$^\circ$ in inverted ordering, respectively~\cite{Esteban:2020cvm}.} in various experiment combinations. It is shown that the combined data of MOMENT, DUNE and T2HKK is sufficient to constrain the $\delta_{CP}$ below 12$^\circ$ at 1$\,\sigma$ CL. This resolution is maintained regardless of the true value of $\delta_{CP}$ or the neutrino mass ordering. In contrast, MOMENT can be anticipated to measure $\delta_{CP}$ at approximately 8$^\circ$...18$^\circ$ precision on its own. MOMENT is therefore expected to have promising prospects in constraining the value of $\delta_{CP}$. 

\section{Summary and outlook}
\label{sec:concl}

\noindent The near-future of neutrino oscillation physics can be anticipated to show advances in the precision measurements of standard neutrino oscillation parameters. In this proceeding, we have briefly reviewed the prospects to determine the value of the Dirac {\em CP} in the proposed medium-baseline muon-decay neutrino experiment MOMENT. Owing to the low beam-related backgrounds, greater selection of oscillation channels and its well-understood neutrino source, the baseline configuration of MOMENT has been shown to be able to provide competitive sensitivities to $\delta_{CP}$. As an independent experiment, MOMENT is expected to reach about 8$^\circ$...18$^\circ$ precision on $\delta_{CP}$ at 1$\,\sigma$ CL. In combination with the proposed long-baseline experiments DUNE and T2HKK, the precision can be pushed below 12$^\circ$ at 1$\,\sigma$ CL regardless of the true value of $\delta_{CP}$.

\acknowledgments
\noindent This project is partially supported by China Postdoctoral Science Foundation under Grant No.\,2020M672930.

\end{document}